\documentclass{webofc}
\usepackage[varg]{txfonts}   
\usepackage{csquotes}
\usepackage{amsmath}

\usepackage{subfigure}
\begin{document}
\title{Deconfinement in pure gauge SU(3) Yang-Mills theory: the ghost propagator}
%
%

\author{\firstname{Orlando} \lastname{Oliveira}\inst{1}\fnsep\thanks{\email{orlando@uc.pt}} \and
        \firstname{Vítor} \lastname{Paiva}\inst{1}\fnsep\thanks{\email{vpaiva462@gmail.com}} \and
        \firstname{Paulo} \lastname{Silva}\inst{1}\fnsep\thanks{\email{psilva@uc.pt}}
}

\institute{ CFisUC, Department of Physics, University of Coimbra, 3004-516 Coimbra, Portugal}

\abstract{%
  The ghost propagator in Landau gauge is studied at finite temperature below and above $T_c$ using lattice QCD simulations. For high temperatures, we find that the ghost propagator is enhanced, compared to the confined phase. The results suggest that the ghost propagator can be used to identify the phase transition, similarly to the gluon propagator case.
}
\maketitle
\section{Introduction}
\label{intro}
The QCD phase diagram has been the subject of several recent theoretical studies, motivated by heavy ion experimental programs. At zero density, one expects a phase transition where quarks and gluons become deconfined at high temperatures. The Polyakov loop $L$ is the order parameter for this transition: for temperatures below the critical temperature $T_c$, $L=0$ and quarks and gluons are confined inside hadrons. For pure gauge theories $T_c=270$ MeV; the inclusion of dynamical quarks lowers this value to $T_c\sim170$ MeV.

In QCD, propagators of fundamental fields encode information about non-perturbative phenomena, such as confinement, deconfinement and chiral symmetry breaking. Following our previous studies of the Landau gauge gluon \cite{Silva:2013maa, Silva:2016onh} and quark \cite{Oliveira:2019erx,Silva:2019cci} propagators at finite temperature, here we study the behaviour of the ghost propagator in Landau gauge at finite temperature. 
\section{Ghost Propagator}
\label{sec-2}
\subsection{Setup}
\label{subsec-2.1}

On the lattice, the computation of the ghost propagator relies on the inversion of a discretized version of the Faddeev-Popov matrix. For details see, for example, \cite{Cucchieri:2018leo}.

In order to evaluate the behaviour of the ghost propagator below and above the critical temperature, a number of lattice ensembles were considered, covering a range of temperatures from 121 MeV up to 486 MeV, as summarized in table \ref{tab-1}, where  $L_s$ is the number of lattice sites in any spatial direction, $L_t$ is the number of lattice sites in the temporal direction and $a$ is the lattice spacing. The temperature is defined as $T=1/(a L_t)$. Following previous works, here we only consider the first Matsubara frequency.

\begin{table}[h]
\centering
\caption{Lattice setup.}
\label{tab-1}
\begin{tabular}{cccccc}
\hline
Temp. (MeV) &    $\beta$ & $L_s$ &  $L_t$ & a [fm] & 1/a (GeV) \\
\hline
121 &    6.0000 & 64    &     16 &     0.1016 &     1.943 \\
194 &    6.0000 & 64    &     10 &     0.1016 &     1.943 \\
243 &    6.0000 & 64    &     8 &     0.1016 &     1.943 \\
260 &    6.0347 & 68    &     8 &     0.09502 &     2.0767 \\
265 &    5.8876 & 52    &     6 &     0.1243 &     1.5881 \\
275 &    6.0684 & 72    &     8 &     0.08974 &     2.1989 \\
324 &    6.0000 & 64    &     6 &     0.1016  &      1.943 \\
366 &    6.0684 & 72    &     6 &     0.08974 &     2.1989 \\
486 &    6.0000 & 64    &     4 &     0.1016  &      1.943 \\
\hline
\end{tabular}
\end{table}


For each of the temperatures studied, we used a lattice ensemble of 100 configurations. Since an \enquote{all-to-all} propagator would be computationally extremely costly, two point sources are considered for each configuration, one at the origin of the lattice, (0, 0, 0, 0), and one at the lattice's spatial midpoint, ($L_s$/2, $L_s$/2, $L_s$/2, 0). A simple average over the two is taken in order to mimic an \enquote{all-to-all} propagator with \enquote{point-to-all} propagators.

In order to account for lattice artefacts for large momenta, the (physical) momenta above 1 GeV were subject to a cylindrical cut \cite{Leinweber:1998uu} where only momenta whose distance, $d$, from the lattice's diagonal was such that $d\, a < 4 \,(2\pi/L_s)$ were considered in the final data -- that is, momenta less than four spatial units away from the lattice's diagonal, ($p$, $p$, $p$, 0).

The propagators pertaining to different temperatures were renormalized at $\mu=4$ GeV, by imposing $G(\mu^2)=1/\mu^2$. In order to do so, a fit was performed to the propagators, with the functional form
\begin{equation}
    \label{eqn-1}
    G(p^2)=\frac{b+cp^2}{p^4+dp^2+e} \hspace{3mm},
\end{equation}
 where $b$, $c$, $d$ and $e$ are adjustable parameters.

\subsection{Temperature Dependence}
\label{subsec - 2.2}

\begin{figure}[h]
\centering
\includegraphics[angle=270, width=8cm]{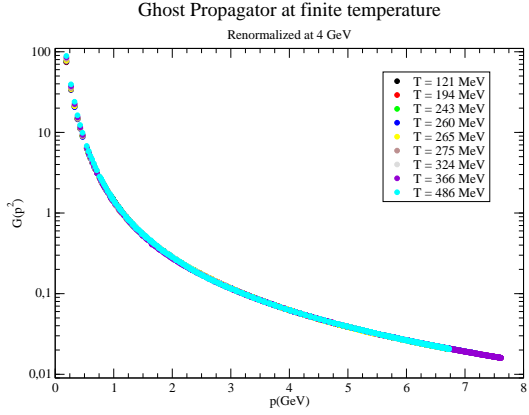}
\caption{Renormalized ghost propagator at finite temperature.}
\label{fig-1}       
\end{figure}

\begin{figure}[h]
\centering
\includegraphics[angle=270, width=8cm]{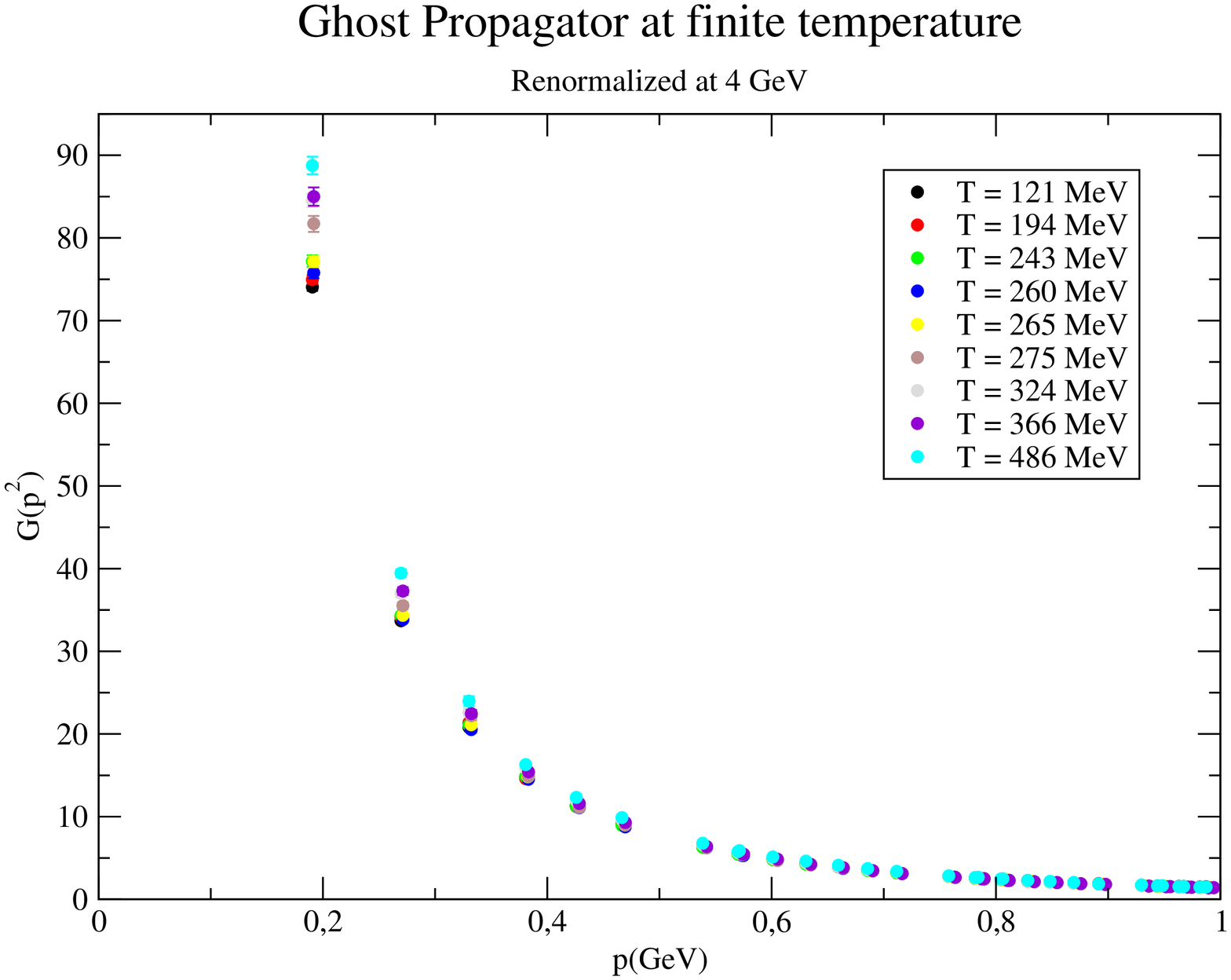}
\caption{Renormalized ghost propagator at finite temperature in the IR region.}
\label{fig-2}       
\end{figure}

The effect of temperature in the ghost propagator for all momentum range is exhibited in figures \ref{fig-1} and \ref{fig-2}. Note that our results are similar to previous results using quenched ensembles with smaller lattice volumes \cite{PhysRevD.85.034501}.

\begin{figure}
    \centering
    \subfigure{\includegraphics[angle=270, width=0.45\textwidth]{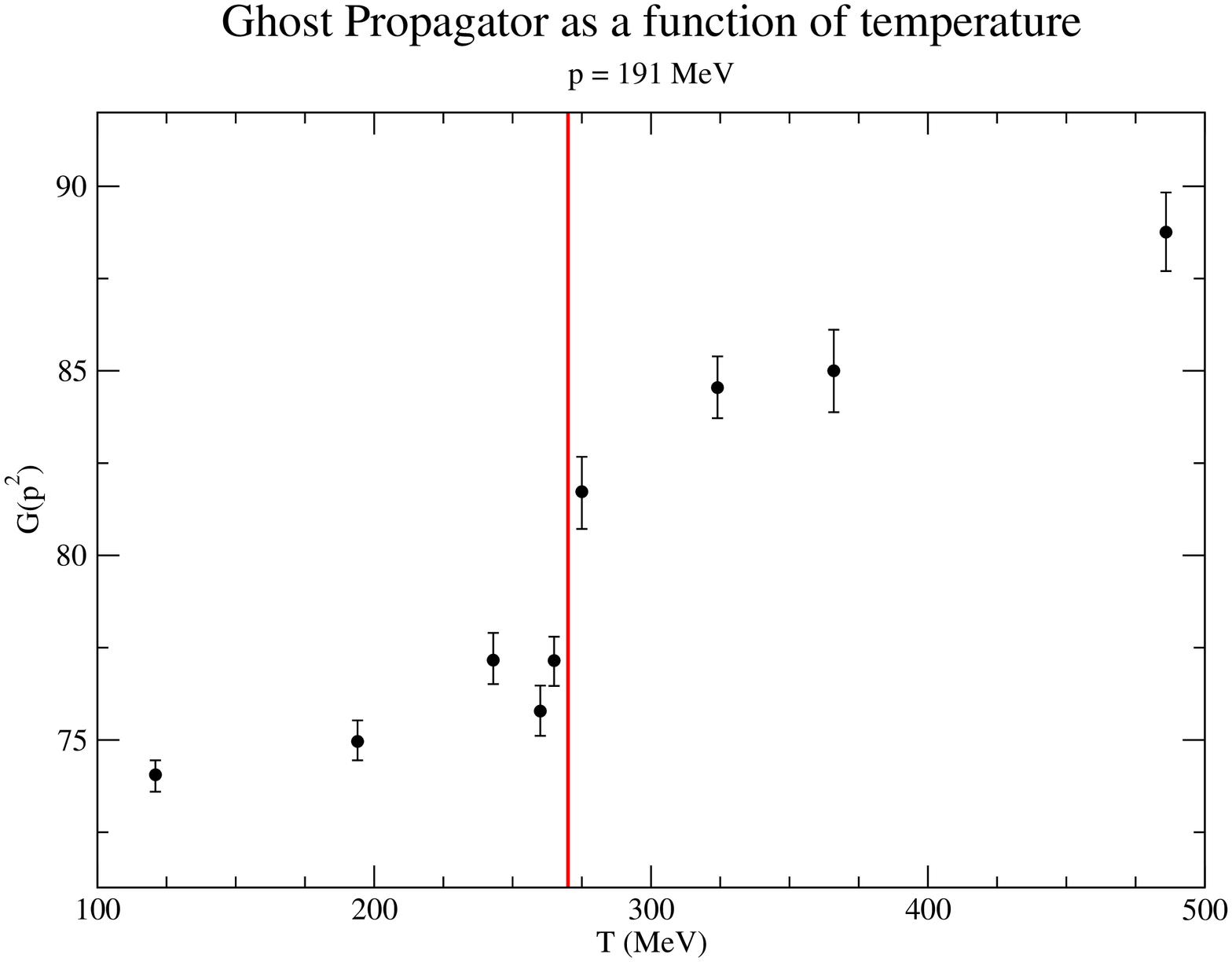}} 
    \subfigure{\includegraphics[angle=270, width=0.45\textwidth]{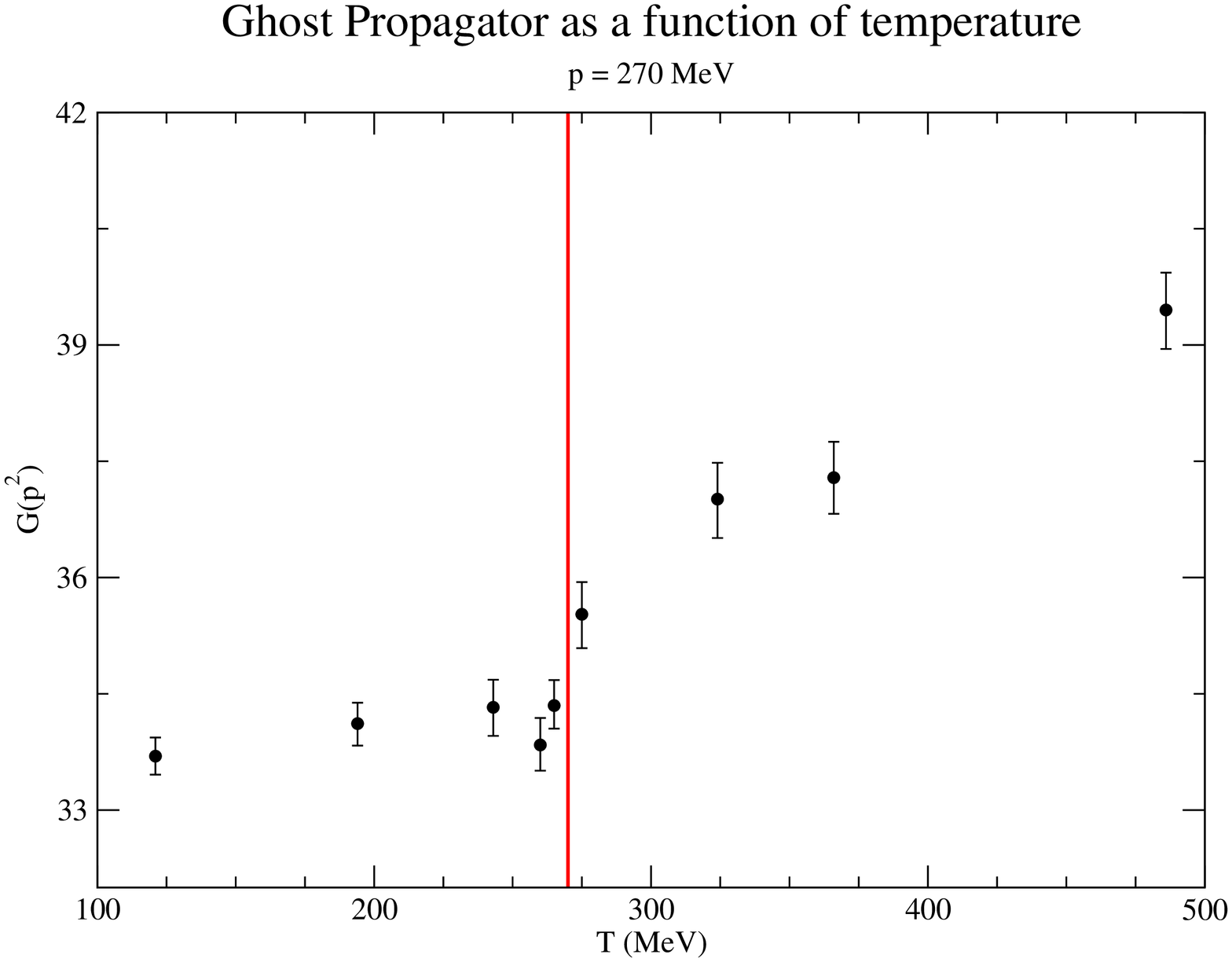}} 
    \subfigure{\includegraphics[angle=270, width=0.45\textwidth]{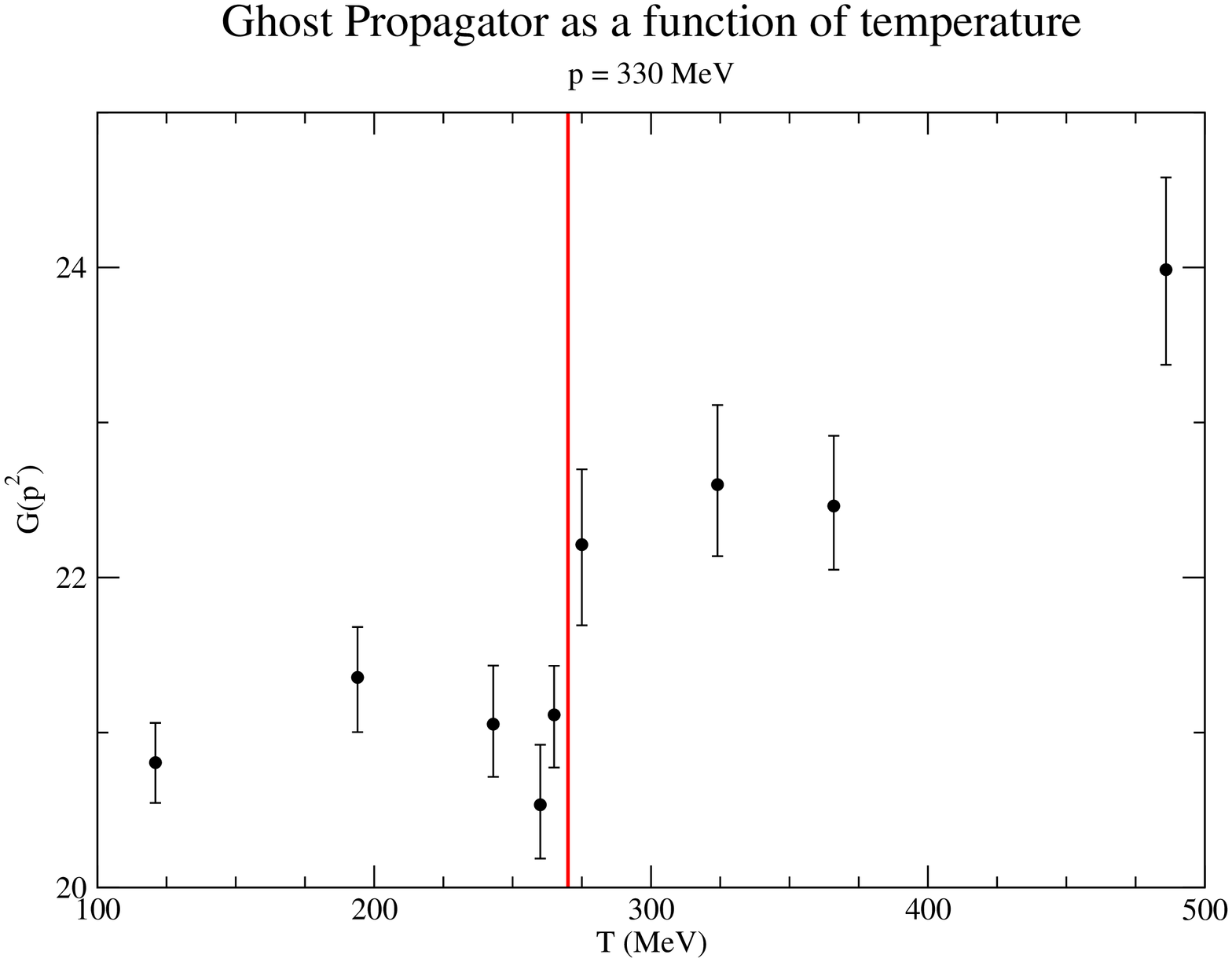}}
    \subfigure{\includegraphics[angle=270, width=0.45\textwidth]{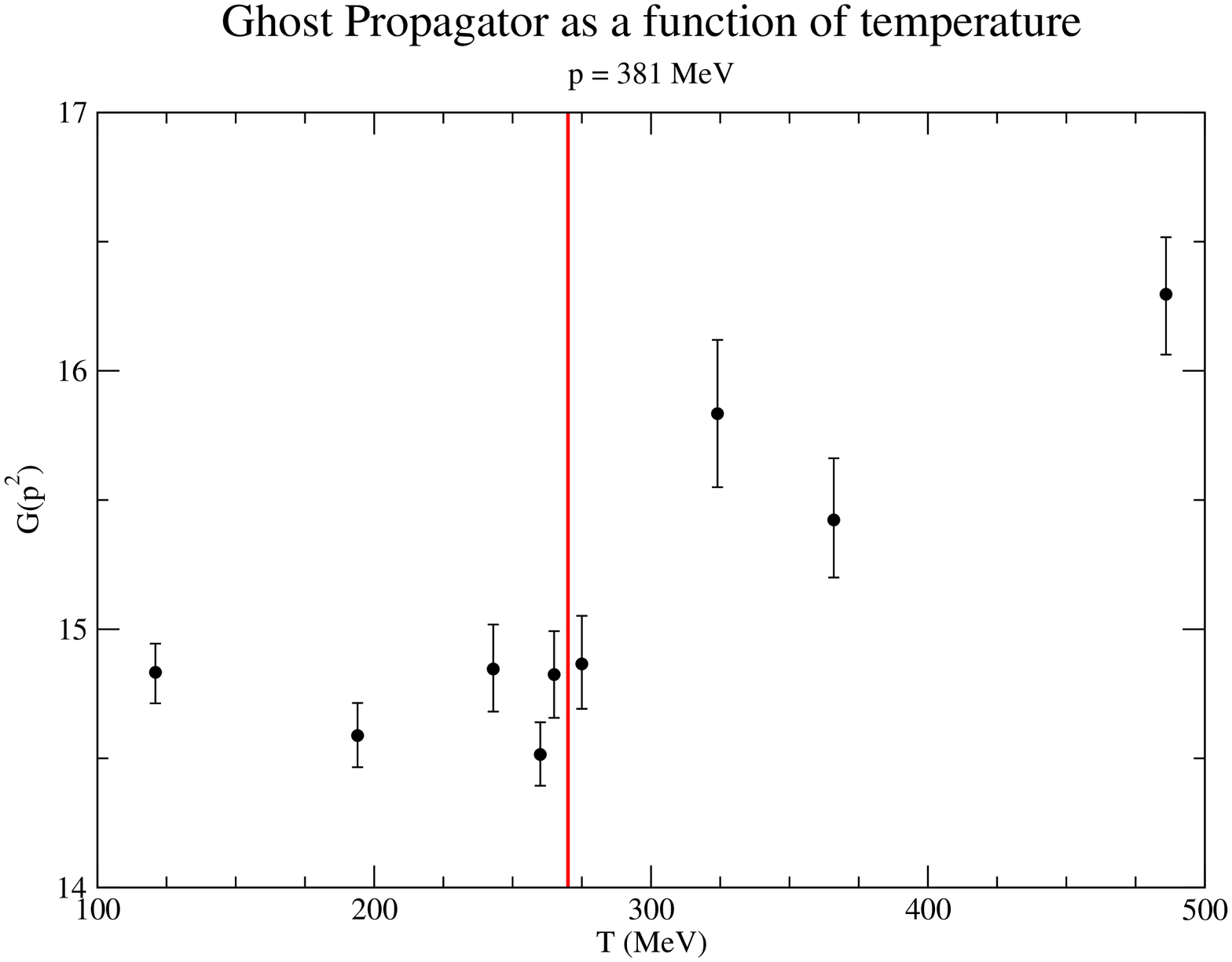}}
    \caption{Ghost propagator as a function of temperature for $p=191$ MeV (top left panel), $p=270$ MeV (top right panel), $p=330$ MeV (left bottom panel) and $p=381$ MeV (right bottom panel). The red vertical line indicates the critical temperature $T_c$.}
    \label{fig-3}
\end{figure}

The distinction between the behaviour below and above the critical temperature is only made clear at lower values of the momenta, as was also observed for the gluon propagator. Figure \ref{fig-2} zooms in on the infrared (IR) region of the ghost propagator, where the enhancement of the propagator above $T_c$, relative to the confined case, is visible. Below the critical temperature, the propagators for the different temperatures are compatible within statistical errors. As Figure \ref{fig-3} further illustrates for the four lowest accessible momenta, the enhancement effect rapidly decreases as the momentum increases and the two regimes become indistinguishable for high momenta.

\subsection{$Z_3$ Dependence}
\label{subsec - 2.3}

On the lattice, gauge configurations related to each other through a center (or $Z_3$) transformation are equivalent. The Wilson gauge action is invariant under a center transformation, which consists in the multiplication of all time links in a constant temporal hyperplane, $x_4=const$, by an element $z$ of the center (or $Z_3$) group,
\begin{equation}
    \label{eqn-2}
    Z_3 = \{e^{-i \frac{2\pi}{3}}, 1, e^{i \frac{2\pi}{3}}\} \hspace{2mm}.
\end{equation}
The symmetry holds for closed loops like the Wilson loop. The Polyakov loop, $L(\Vec{x})$, however, is not invariant under such a transformation, $L(\Vec{x}) \rightarrow zL(\Vec{x})$. It thus constitutes an order parameter for the deconfinement phase transition. Below $T_c$, center symmetry holds and $\langle L \rangle = 0$; above $T_c$, center symmetry is spontaneously broken, the $Z_3$ sectors are not equally populated and $\langle L \rangle \neq 0$.

\begin{figure}
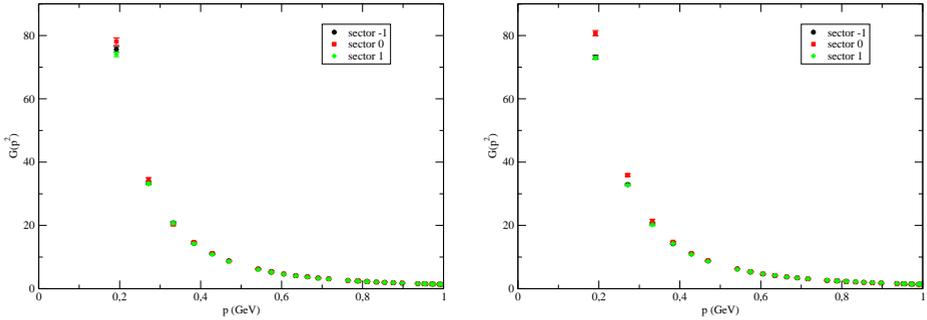

    \centering
    \subfigure{\includegraphics[width=0.45\textwidth]{ghostZ3belowTc.eps}}
    \hspace*{0.3cm}
    \subfigure{\includegraphics[width=0.45\textwidth]{ghostZ3aboveTc.eps}} 
    \caption{Ghost propagator's sector dependence below $T_c$ (left-hand panel at $T=270$ MeV) and above $T_c$ (right-hand panel at $T=274$ MeV).}
    \label{fig-4}
\end{figure}

Previous works have shown that the gluon \cite{Silva:2016onh} and quark propagators \cite{Silva:2019cci} are sensitive to the $Z_3$ sector of the gauge configurations. Our preliminary results suggest that the ghost propagator is also  sensitive to the $Z_3$ sector above $T_c$. Figure \ref{fig-4} shows the IR region of two lattice simulations  with $L_s=72$ and $L_t=8$ with $\beta=6.058$ (left-hand panel) and $\beta=6.066$ (right-hand panel). The results show that the ghost propagator behaves differently below and above $T_c$, with a suppression of the $\pm 1$ sectors relative to the $0$ sector for the deconfined phase. As we found previously for the gluon propagator \cite{Silva:2016onh}, the $\pm 1$ sectors are indistinguishable above $T_c$.

\section{Conclusions and outlook}

In this paper we study the Landau gauge ghost propagator at finite temperature using lattice simulations. We found an enhancement of the ghost form factor above the critical tempera\-ture $T_c$, already found in previous SU(3) studies on smaller volumes \cite{PhysRevD.85.034501}. Note that early SU(2) studies concluded in favour of a nearly independent ghost propagator with the temperature \cite{Cucchieri:2007ta}. We also show preliminary results for the $Z_3$ dependence of the ghost propagator. Although the propagators in the various sectors are indistinguishable below $T_c$, we found a suppression, above $T_c$, of the $\pm 1$ sectors in comparison with the $0$ sector. However, in the deconfined phase  the $\pm 1$ sectors are still compatible within errors.

We are currently extending the study of the $Z_3$ dependence for other temperatures. In the near future we also plan to study the QCD propagators at finite temperature using dynamical configurations.

\section*{Acknowledgements}

This work was partly supported by the FCT – Fundação para a Ciência e a Tecnologia, I.P., under Projects Nos. UIDB/04564/2020,  UIDP/04564/2020 and CERN/FIS-COM/0029/2017. P. J. S. acknowledges financial support from FCT (Portugal) under Contract No. CEECIND/00488/2017. The authors acknowledge the Laboratory for Advanced Computing at the University of Coimbra (http://www.uc.pt/lca) for providing access to the HPC resource Navigator.

\bibliography{references}

\end{document}